# Thermal regimes of HTS cylinders operating in devices for fault current limitation

V Meerovich and V Sokolovsky

Physics Department, Ben-Gurion University, Beer-Sheva 84105, Israel

E-mail: **victorm@bgu.ac.il**

**Abstract.** We reveal obstacles related to the application of HTS cylinders in current limiting devices based on the superconducting – normal state transition. It is shown that, at the critical current density achieved presently in bulk materials, and especially in BSCCO-2212, the required thickness of the cylinder wall in a full-scale inductive device achieves several centimeters. A simple mathematical model of the operation of an inductive fault current limiter (FCL) is used to show that such cylinders cannot be cooled in admissible time after a fault clearing and, hence, the inductive FCLs and current-limiting transformers employing BSCCO cylinders do not return to the normal operation in the time required. For the recovery even with a non-current pause in the circuit, the cylinders are needed with the critical current density by an order higher than the existed ones.

## 1. Introduction

Despite more than 30 years history of the development of superconducting (SC) fault current limiters (FCLs) and other devices based on the superconducting-normal state transition (S-N transition), number of papers and projects devoted to the topic are not exhausted [1-6]. The first prototypes of SC FCLs were based on low-temperature superconductors (LTS) [7-9]. One popular and widely studied design was an inductive device, consisting of two magnetically coupled windings: a primary winding inserted in series in a power circuit and a secondary SC short-circuited winding [7, 8, 10]. The S-N transition in the SC winding or in its part causes the increase of the resistance of this winding resulting in the increase of the FCL impedance. This design has some excellent features [10]: (a) absence of electric connections between the superconducting elements and the circuit to be protected; (b) physical separation of cryogenic and normal parts of the device, i. e. no current leads into a cooled zone are needed, etc.

In the designs based on LTS materials [7, 8], the secondary coil was fabricated from a well-stabilized superconducting wire and short-circuited by a special superconducting switching element. Only the switching element undertook the S-N transition under a fault while the secondary coil remained in the superconducting state. The same design was also considered for high-temperature superconducting (HTS) inductive FCLs and current-limiting transformers [8, 11, 12].

The progress in the fabrication technology of large diameter (40-50 cm) HTS rings and hollow cylinders [13, 14] stimulated their application as secondary one-turn short-circuited SC windings of the inductive FCL. Based on these cylinders, different experimental models of FCLs [3, 6, 15-17] and current-limiting transformers [5, 18] were built with the purpose to prove the feasibility of concepts and to study the device operation in power systems.

It is known that the operation of an FCL in a power circuit specifies clear requirements to the thermal regime of an element undertaking the S-N transition in the device [10]. This element should be in the stable SC state below than double amplitude of the nominal current and quickly (during 0.003-0.01 s) pass into the normal state when the instantaneous circuit current exceeds a definite value (so called "activation current", $I_{act}$). It should return into the SC state in a short time (about 1s) after a fault clearing, i. e. at the time of automatic reclosing of the damaged line. The last is required to keep the static and transient stability of the parallel operation of electric machines in a power system and to limit the losses in the FCL [10, 19]. Most of studies performed on small models have concentrated attention on the peculiarities of the S-N transition [3, 5, 17, 18, 20-22]. The topic of a quick recovery of the SC state in the cylinders was presented more modest. There are several publications which report a fast return of small laboratory models into the initial state [23-25]. Other papers note that even small power models demonstrate a sufficient delay of the recovery after the operation [26, 27]. With the increase of the power of tested prototypes, the issue of the recovery is almost completely disappears from publications.

Simple estimates performed by us for full-scale FCLs and current-limiting transformers have shown that the recovery problem become sharper with the increase of the device power [28]. In the present paper we analyze this problem in more detail.

## 2. Mathematical model for a SC cylinder with uniform temperature distribution

In order to show how the thermal processes in a superconducting cylinder influence the behavior of a full-scale inductive FCL, let us consider a simple mathematical model of the FCL operation [10]. The model is based on the equations for the equivalent circuit of a two-winding transformer inserted in series in a single-phase circuit (Fig. 1):

$$U_{10}\sin(\omega t) = R\,i + (L_{\sigma 1} + L)\frac{di}{dt} + L_{\sigma 2}\frac{di_s}{dt} + u_s;$$

$$u_s + L_{\sigma 2}\frac{di_s}{dt} = L_n\frac{di_1}{dt};  \qquad (1)$$

$i = i_s + i_1$

combined with the non-stationary heat equation for a homogeneous superconductor:

$$c_V S\Delta \frac{dT}{dt} = u_s i_s - hS(T - T_0). \qquad (2)$$

Here $U_{10}$ and $\omega$ is the amplitude and cyclic frequency of the voltage applied to the circuit; $R$ and $L$ are the resistance and inductance of the circuit which change at a short-circuit from the nominal load values to the short-circuit parameters; $L_{\sigma 1}$ and $L_{\sigma 2}$ are the leakage inductances of the primary and secondary windings; $i$, $i_1$ and $i_s$ are the circuit current, magnetizing current and current in the SC winding, respectively; $L_n$ is the magnetizing inductance; $u_s$ is the active component of the voltage drop across the superconductor; $T$ and $T_0$ are the temperatures of the superconductor and coolant, respectively; $c_V$ is the specific volume heat capacity of the superconductor; $\Delta$ is the thickness of the cylinder wall; $S$ is the cooled surface; $h$ is the heat transfer coefficient from a wall of the superconductor in coolant; $t$ is the time. All the parameters are referred to the primary. The superconductor is approximated by a non-linear voltage-current characteristic including three portions corresponding to the flux creep, flux flow and normal state [28-30]:

$$u_s = \begin{cases} u_{s0}\left(\dfrac{|i_s|}{I_c(T)}\right)^n \left(1-\dfrac{T-T_0}{T_c-T_0}\right)\text{sign}(i_s) & \text{if } |i_s| \leq I_b(T) \text{ and } T < T_c \\ u_{s0}\left(\dfrac{I_b(T)}{I_c(T)}\right)^n \left(1-\dfrac{T-T_0}{T_c-T_0}\right)\text{sign}(i_s) + R_n[i_s - \text{sign}(i_s)I_c(T)] & \text{if } |i_s| > I_b(T) \text{ and } T < T_c \\ R_n i_s (1+\gamma(T-T_0)) & T \geq T_c \end{cases} \quad (3)$$

where $u_{s0}$ is the voltage drop in the cylinder at the critical current $I_c$ determined from the criterion of 1μV/cm; $I_b(T)$ is the boundary current value separating the zones of the flux creep and flux flow; $T_c$ is the critical temperature; $R_n$ is the normal state resistance and γ is its temperature coefficient.

Here we assume the linear dependence of $I_c$ on temperature:

$$I_c(T) = I_{c0}\left(1 - \frac{T-T_0}{T_c-T_0}\right). \qquad (4)$$

In the case of employing of a superconducting cylinder as the secondary, the resistance in the normal state is determined by

$$R_n = \frac{\rho_n \pi D}{l\Delta} \qquad (5)$$

where $\rho_n$ is the normal state resistivity; $D$ and $l$ are the diameter (which is approximately equaled to the diameter of the magnetic core) and height of the cylinder, respectively.

The cross-section area $l\Delta$ of the cylinder equals to the ratio of its critical current $I_c$ to the critical current density $J_c$. The required critical current is calculated as $I_c = I_{act} w_1$ where $w_1$ is the turn number in the primary winding. The required wall thickness of the superconducting cylinder is obtained as

$$\Delta = \frac{I_{act} w_1}{l J_c}, \qquad (6)$$

and it is one from the main parameters determining the rate of heating and cooling, as well as the losses in the cylinder in the non-SC state.

For a frequently considered open core design of the inductive FCL, the turn number $w_1$ and core diameter $D_{core} \sim D$ are determined by simple expressions [27]:

$$w_1 = K_2(\omega L_n/I_M)^{1/3}\,;\quad D_{core} = K_1(\omega L_n I_M^2)^{1/3}, \qquad (7)$$

where $I_M$ are the effective magnetizing current; $K_1$ and $K_2$ are the coefficients determined by the design features and admissible maximum magnetic flux density $B_{max}$ in the core ($K_1 = 0.00219$, $K_2 = 937$ for the distribution voltage class).

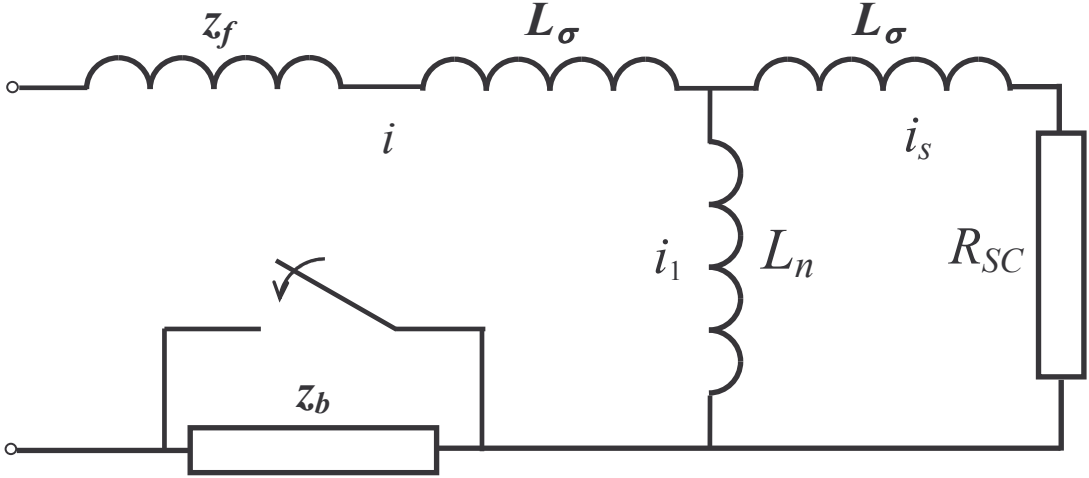

Fig. 1. Equivalent circuit of a circuit with an inductive FCL: $z_f$ and $z_b$ are the short-circuit and load impedances, respectively; $L_{\sigma 1}$ and $L_{\sigma 2}$ are the leakage reactances of the primary and secondary windings of the FCL; $L_n$ is the magnetizing inductance of the FCL; $R_{SC}$ is the non-linear resistance of the SC winding. The switch short-circuited the load simulates a fault.

## 3. Performance of a full-scale inductive FCL

Let us consider an inductive FCL installed in each phase of a three-phase 35/11 kV, 45 MVA transformer from the low-voltage side, i. e. into a circuit with the phase voltage of 6.3 kV and rated current of 2400 A.

Suppose the parameters of the cylinder: the normal state resistivity $\rho_n = 10^{-5}$ $\Omega \cdot m$, critical temperature $T_c = 105$ K, and specific volume heat capacity $c_V = 0.7 \times 10^6$ J/(m³K). These are typical parameters for BSCCO-2212 cylinders fabricated for inductive FCLs [20, 31]. It is assumed that cooling is provided by liquid nitrogen at $T_0 = 77$ K, and the heat transfer coefficient to coolant is $5 \times 10^3$ W/m²K.

Fig. 2 presents the required wall thickness as a function of the normalized limited current $I_L$ which could be achieved at a complete transition of the cylinder into the normal state. The calculations were performed for the minimum acceptable activation current [10] $I_{act} = 2\sqrt{2}I_{nom}$, $l = 1$ m and different values of $J_c$.

The critical current density for bulk MCP-BSCCO 2212 cylinders does not exceed $2\times10^7$ A/m$^2$ at the temperature of 77 K [31]. For this density, the limitation of the current to $2I_{nom}$ requires a 2 cm cylinder. To achieve a deeper limitation, one should increase the impedance of the primary winding and, according to (7), the turn number $w_1$ (see (6)). It leads to the rise of the required thickness.

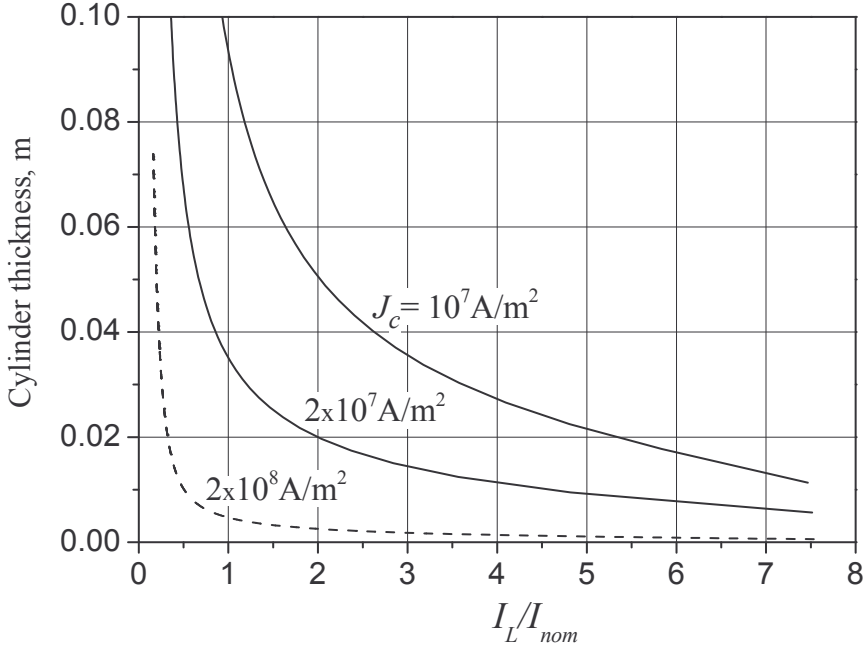

Fig. 2. Thickness of the cylinder wall as a function of the normalized limited current for different critical current densities.

Fig. 3 shows the change of temperature of a BSCCO cylinder during a fault and after a fault clearing for the limited current $I_L = 2I_{nom}$ and $J_c = 2\times10^7$ A/m$^2$. The main parameters of the FCL are $D = 0.62$ m, $w_1 = 56$. Two typical regimes of a power circuit before the automatic reclosing were considered: 1) after a fault clearing by a circuit-breaker the circuit operates with the nominal load (the FCL before the buses, a fault in one of the lines); 2) after a fault clearing the circuit is disconnected (1 s non-current pause, the FCL in a damaged line). The results show that, even with a non-current pause, the cylinder is not cooled sufficiently: the automatic reclosing of the circuit in about 1 s leads to the FCL activation in the nominal regime of the circuit.

Fig. 4 illustrates how the results could be changed if we had at our disposal the cylinder with the critical current density of $2\times10^8$ A/m$^2$, i. e. ten times larger than the existed one. Even in this case, a non-current pause is required to return the cylinder into the initial state. A steady-state regime with the dissipation (curve 1) has been observed in experiments even with small-scale models [26].

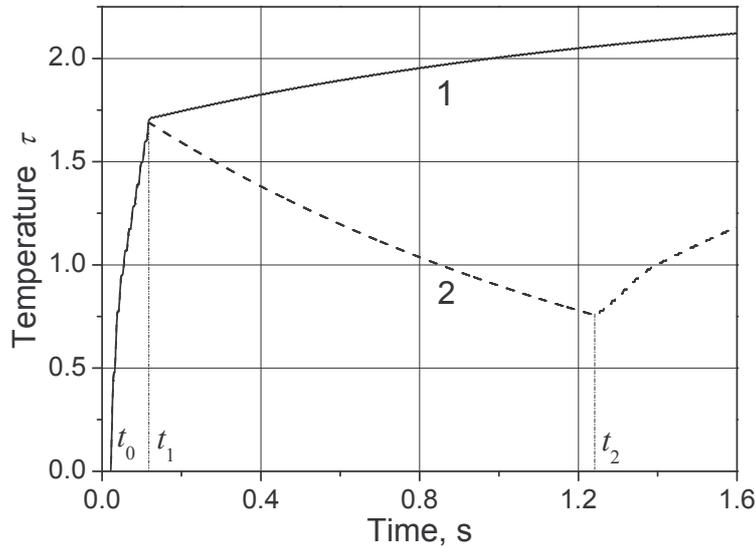

Fig. 3. Temperature of a BSCCO-2212 cylinder with $J_c = 2\times10^7$ A/m$^2$ at the operation of the FCL. 1- with a non-current pause; 2 - without a pause. The characteristic time points: $t_0$ is the time of a fault; $t_1$ is the fault clearing; $t_2$ is the reclosing of the line after a non-current pause. Here and below $\tau = \dfrac{T-T_0}{T_c-T_0}$ is the normalized temperature.

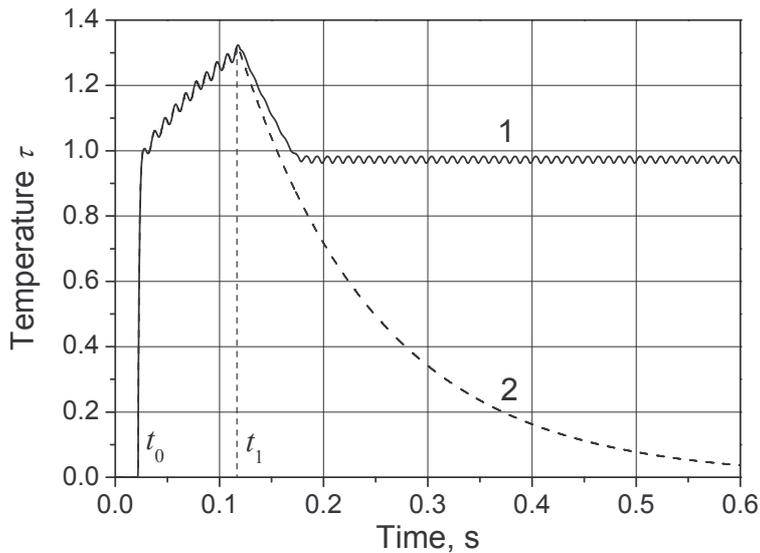

Fig. 4. Temperature of a cylinder with $J_c = 2\times10^8$ A/m$^2$ at the operation of the FCL. 1- with a non-current pause; 2 - without a pause.

## 4. Account of non-uniform temperature distribution in a cylinder wall

One can assume that, increasing the ratio $I_L/I_{nom}$ and thus decreasing the wall thickness (Fig. 2), one can achieve the recovery of a BSCCO cylinder though with a non-current pause. To analyze this possibility in more detail, let us consider the cooling during a non-current pause with taking into account a low heat conductivity of HTS materials (for BSCCO-2212, it is of the order of 1 W/m·K at 77 K [32]).

The temperature distribution in the wall of a cylinder with $\Delta \ll D$ is described by the following equation

$$c_V \frac{\partial T}{\partial t} = \lambda \frac{\partial^2 T}{\partial x^2} \tag{8}$$

with the boundary condition

$$\lambda \frac{\partial T}{\partial x} = h(T - T_0) \quad \text{at } x = 0 \text{ and} \quad \lambda \frac{\partial T}{\partial x} = -h(T - T_0) \quad \text{at } x = \Delta.$$

The typical temperature distributions in the walls cooled from both surfaces is presented in Fig. 5. The cylinders are assumed to be initially heated up to the critical temperature ($\tau = \frac{T - T_0}{T_c - T_0} = 1$). Due to the linearity of the heat equation, the temperature distribution is proportional to the initial temperature of the heating.

One can see that only a small surface layer of a thick cylinder can be cooled down in 1 s. The non-uniform temperature distribution leads to inhomogeneity of the critical current density $J_c$ over the specimen. The critical current at any instant is determined by integrating $J_c$ over the specimen thickness. Assuming the linear temperature dependence of $J_c$ as in (4), we have calculated the critical current of a cylinder after 1 s. Even in the best case, when the cylinder is heated only up to the critical temperature, the critical current is 12% of the critical current at 77 K for the wall thickness of 2 cm and 50% for the wall thickness of 0.5 cm. Only for 0.2 cm thickness, the critical current is recovered almost completely during 1 s pause. However, with increasing the heating temperature during the current limitation, the reduction of the critical current becomes noticeable even for a 0.2 cm cylinder.

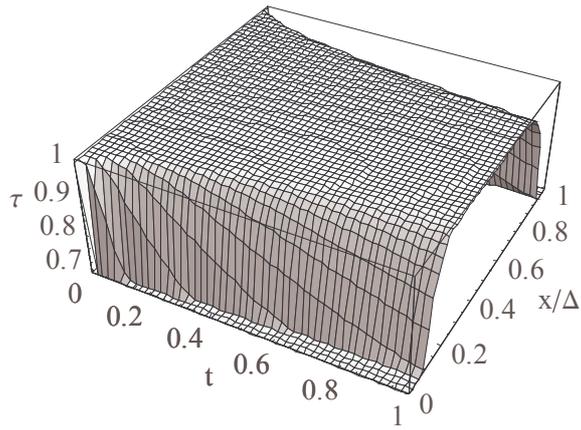

(a)

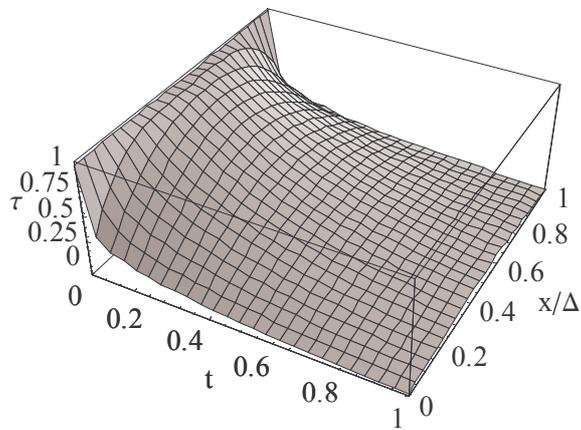

(b)

Fig. 5. Evolution of the temperature distribution in the cylinder wall heated initially to the critical temperature. The wall thickness $\Delta$ is 2 cm (a) and 0.2 cm (b).

## 5. Discussion

Let us make some remarks on the obtained results.

1. The heat transfer coefficient chosen for our simulations is beyond its real value. It is known that when the temperature of a cooled surface exceeds the temperature of liquid nitrogen by 10-15 K, the transition happens from the nuclei to the film boiling (boiling crisis). At the boiling crisis, the heat transfer coefficient falls off by a factor of about 10. A cylinder heated up to the critical temperature is under the conditions of the film boiling. Thus, the obtained rate of the cooling is even too optimistic.

2. Let us evaluate the maximum power of a 10 kV circuit at which a BSCCO cylinder is recovered within 1 sec non-current pause. To provide the recovery, the wall thickness has to be reduced to 2 mm, i. e. ten times in the comparison with the example above considered (Fig. 5). Such a cylinder is applicable for $I_{act}$ = 480 A (see Eq. 6) and, hence, for maximum current $I_{nom}$ =240 A, i. e. for the power 2.4 MVA.

3. Use of the closed magnetic system instead of the open core does not reduce the turn number in the primary winding [10, 27] and, hence, requires the same thickness of the cylinder wall.

4. A natural way to decrease the required thickness for a higher power is to increase $J_c$. The rise of $J_c$ by an order of magnitude could allow to use the cylinders with the thickness 7-8 times less (see the dotted curve in Fig. 2). Note, that the value of the critical current density in melt-casting BSCCO-2212 cylinders and tubes was practically unchanged in the past decade [31] and seemingly close to its limit. At the same time, the critical current was markedly increased in melt-processed YBCO products. However, inductive FCLs and current-limiting transformers using YBCO cylinders with a relatively high $J_c$ also require a non-current pause for the recovery.

We conclude, that, in full-scale inductive devices, BSCCO cylinders do not return to the normal operation in a time required. Therefore, presently fabricated BSCCO-2212 cylinders cannot be used in inductive FCLs designed for industrial application. The cylinders are inapplicable in current-limiting transformers because, as simple calculations show, a 15 MVA current-limiting transformer designed for the deep current limitation [27] is in need of a cylinder of 3.5 cm in thickness. We concede, however, that these cylinders can be used in resistive FCLs shunted by electrical reactors [2], or in some specific systems, which do not require a quick recovery, e. g. at the installation of the inductive FCL in a line with parallel standby lines.

Quick recovery demonstrating in some experiments with small models of the inductive devices can be explained by an incomplete S-N transition in the cylinder. For successful operation of an inductive FCL, the condition $R_{SC} > \omega L_n$ is required (Fig. 1). As the simple estimates show, the resistance appeared at the S-N transition of a section of 0.1 mm in length can be sufficient for the operation of a small inductive device [20, 33]. This section can be cooled sufficiently quickly because the heat is not only removed from the surface but also propagates to nearby sections. However, to fulfill the condition $R_{SC} > \omega L_n$ in a large power device, the complete circumference of the cylinder has to pass into the normal state [16, 27].

One possible solution of the recovery problem in SC inductive devices could be the use of special switching elements closing the SC winding [7, 8, 11, 34]. These elements can be performed with coated conductors or with thin films.